**Perfect absorption by an atomically thin crystal**

Jason Horng[1,2][†]*, Eric W. Martin[1][†], Yu-Hsun Chou[1,3], Emmanuel Courtade[4], Tsu-chi Chang[3], Chu-Yuan Hsu[3], Michael-Henr Wentzel[1], Hanna G. Ruth[1], Tien-chang Lu[3], Steven T. Cundiff[1], Feng Wang[2], Hui Deng[1]*

[1] Department of Physics, University of Michigan, Ann Arbor, Michigan 48109-1040, United States

[2] Department of Physics, University of California at Berkeley, Berkeley, California 94720, United States.

[3] Department of Photonics, College of Electrical and Computer Engineering, National Chiao Tung University, Hsinchu 300, Taiwan

[4] Université de Toulouse, INSA-CNRS-UPS, LPCNO, 135 Av. Rangueil, 31077, Toulouse, France

[†]These authors contributed equally to this work

*To whom correspondence should be addressed. Email: dengh@umich.edu, jahorng@umich.edu

**Abstract**

Optical absorption is one of fundamental light-matter interactions. In most materials, optical absorption is a weak perturbation to the light. In this regime, absorption and emission are irreversible, incoherent processes due to strong damping. Excitons in monolayer transition metal dichalcogenides, however, interact strongly with light, leading to optical absorption in the non-perturbative regime where coherent re-emission of the light has to be considered. Between the incoherent and coherent limits, we show that a robust critical coupling condition exists, leading to perfect optical absorption. Up to 99.6% absorption is measured in a sub-nanometer thick $MoSe_2$ monolayer placed in front of a mirror. The perfect absorption is controlled by tuning the exciton-phonon, exciton-exciton, and exciton-photon interactions by temperature, pulsed laser excitation, and a movable mirror, respectively. Our work suggests unprecedented opportunities for engineering exciton-light interactions using two-dimensional atomically thin crystals, enabling novel photonic applications including ultrafast light modulators and sensitive optical sensing.

**Introduction**

Optical absorption in semiconductors results from excitation of electrons from the valence to conduction band, followed by incoherent dissipation. The absorption is greatly enhanced at the exciton resonances due to the coherent nature of the exciton dipole[1,2], up to a few percent in a single gallium arsenide(GaAs) quantum well of a few nanometers in thickness[3,4]. Nevertheless, excitonic absorption remains an incoherent and perturbative process in typical low-dimensional semiconductors without enhancement by high-quality cavities or coherent laser beams[5,6]. In two-dimensional(2D) transition metal dichalcogenide(TMD) monolayers, the exciton transition dipole moment is almost two orders of magnitude stronger than in a GaAs quantum well, resulting from the reduced Coulomb screening in 2D materials[7–9]. High reflectivity has been reported from monolayers at low temperatures[10,11], suggesting that the radiative decay dominates over pure dephasing and inhomogeneous broadening.

Here, by tuning between the incoherent and coherent limits of exciton absorption in a monolayer placed in front of a mirror, we experimentally demonstrate critical coupling between 2D excitons and free space photon, leading to perfect absorption at the exciton resonance by a sub-nanometer thick crystal. Up to 99.6% absorption is measured in a monolayer $MoSe_2$. Furthermore, we show that the critical coupling condition is readily tuned by temperature tuning of the exciton-phonon scattering, by ultrafast laser control of the exciton-exciton scattering, and by tuning of the exciton-photon interactions with a movable mirror. Our work establishes perfect absorption in 2D TMDs using a simple and versatile platform for engineering exciton-light interactions. This work enables novel photonic applications related to coherent perfect absorption (CPA), such as ultrafast light modulators, sensitive photodetection[12,13] and optical sensing[14–16], coherent optical computing[17–19], and novel designs of imaging[20] and laser pulse shaping[21].

When the light-matter coupling is considered self-consistently, the absorption spectrum of a homogeneous thin-film semiconductor, $\alpha(E)$, for a resonance at energy $E_R$ is[22]

$$\alpha(E) = \frac{2\Gamma_{sca}\Gamma_{rad}}{(E-E_R)^2+(\Gamma_{sca}+\Gamma_{rad})^2}, \qquad (1)$$

where $\Gamma_{sca}$ and $\Gamma_{rad}$ are scattering and radiative decay rates, respectively. This absorption can be understood by a simple microscopic model of exciton absorption shown in Figure 1a. In this model, a resonant planewave light field illuminates a homogeneous 2D sheet of dipoles, creating an exciton polarization that either radiatively decays at rate $\Gamma_{rad}$ or scatters at rate $\Gamma_{sca}$. $\Gamma_{rad}$ is directly proportional to the exciton transition dipole moment. It accounts for the coherent re-emission process that returns energy to the field. $\Gamma_{sca}$ includes all mechanisms that scatter the excitons to another state, such as exciton-phonon scattering, exciton-impurity scattering, and non-radiative recombination.

In typical semiconductors, because $\Gamma_{sca} \gg \Gamma_{rad}$ the energy carried by the optically-excited excitons are predominantly transferred into non-radiative decay channels and equation(1) reduces to the Elliott formula for excitonic absorption in the perturbative regime[22]. In this incoherent limit, the total integrated absorption only depends on $\Gamma_{rad}$, which therefore is independent of temperature or excitation power. The constant integrated absorption, $A_0$, is determined by the total number of oscillators in the material as required by the oscillator sum rule.

However, as the radiative decay becomes comparable to or larger than scattering or pure dephasing, re-emission of excitons into the optical field is no longer negligible. When consistently solved using equation (1), the normalized integrated absorption is $A/A_0 = \Gamma_{sca}/(\Gamma_{rad} + \Gamma_{sca})$, the ratio of the exciton loss rate in the material and total rate of exciton decay. The absorption decreases with decreasing $\Gamma_{sca}$ for fixed $\Gamma_{rad}$. At $\Gamma_{sca} \ll \Gamma_{rad}$, $\alpha(\omega)$ vanishes and the excited semiconductor reradiates the entire excitation, which yields 100% reflection at the resonance[10,11]. The reduction of integrated area is, therefore, a direct evidence for the excitonic medium reaching the coherent limit.

The transition from the coherent limit ($\Gamma_{sca} \ll \Gamma_{rad}$) to the incoherent limit ($\Gamma_{sca} \gg \Gamma_{rad}$) is illustrated in Figure 1b for a free-standing excitonic monolayer. In the incoherent limit, the maximum absorption increases with $\Gamma_{sca}$, and therefore the total linewidth is decreased to maintain an approximately constant total absorption (area under the absorption spectrum). In the coherent limit, the maximum absorption decreases while the total linewidth also decreases, leading to decreasing total absorption. The maximum linear absorption peaks at 50% when $\Gamma_{sca} = \Gamma_{rad}$ due to the mirror symmetric of the monolayer[6].

Perfect absorption requires overcoming the 50% maximum absorption, can be achieved by means such as by introducing another laser beam to form a two-port device[17,18,23]. Alternatively, we show that it can also be achieved by breaking the mirror symmetry[14,24] by simply putting a mirror behind the 2D exciton medium(see figure 1d). Critical coupling can be established to achieve perfect absorption due to interference between the reflection and transmission.

To understand the perfect absorption intuitively, we consider the case where resonant light illuminates a monolayer placed one quarter wavelength away from the mirror. In the incoherent limit, the monolayer has high transmission, and most of the incident light is reflected by the mirror, leading to a net reflection coefficient $r = 1$ (referenced to the monolayer). In the coherent limit, the monolayer alone acts as a perfect reflector on resonance, resulting in a reflection coefficient $r = -1$. Since the reflection coefficient at the exciton resonance is real-valued, it must go through zero to transition from the incoherent to coherent limit. That is, a critical coupling condition giving perfect absorption must exist. We show in supplementary I that the perfect absorption condition is robust against inhomogeneous broadening and for arbitrary monolayer-to-mirror distances, since it is required by the continuity of the reflection coefficient.

Quantitatively, the reflection coefficient is obtained by considering all multiple reflections $r = (3r_m + 1)/(1 - r_m)$, where $r_m$ is the reflection coefficient of a free-standing monolayer. The critical coupling condition for perfect absorption is $r_m = -1/3$, or $\Gamma_{sca} = \Gamma_{rad}$, where $\Gamma_{rad}$ is the radiative decay rate of the monolayer in front of a mirror(twice the free-standing monolayer radiative decay rate). Note that this condition also corresponds to integrated $A = A_0/2$.

The critical coupling can also be understood as an analogy to a ring resonator as illustrated in figure1e. The photon channel corresponds to the linear waveguide, and the exciton channel corresponds to the ring waveguide. $\Gamma_{rad}$ and $\Gamma_{sca}$ correspond to the input coupling efficiency and the loss in the ring waveguide, respectively. Similar to the ring resonator, $\Gamma_{rad} = \Gamma_{sca}$ corresponds to a critical coupling condition where all the energy is trapped in the exciton channel and does not re-emit into photons.

To achieve perfect absorption, the coherent limit of exciton absorption must be demonstrated and a knob to tune the system between the two limits is needed. We use a high quality hBN-encapsulated $MoSe_2$ monolayer placed on a distributed Bragg reflector(DBR) with high reflectance. The absorption spectrum, $\alpha(E)$, is obtained directly from $\alpha(E) = 1 - R(E)$, where R is the measured reflection spectrum. Figures 2a and b show an optical image of the sample studied and a schematic of the overall structure. Details of fabrication are described in Methods. Figure 2c shows the position dependence of the electric field from a transfer-matrix calculation (Supplementary II). The monolayer is positioned close to the field maximum, which leads to enhanced radiative decay rate and brings the system closer to the coherent limit.

The coherent effects on the optical absorption are studied by transitioning the system from the incoherent to coherent limit via temperature tuning of the ratio $\Gamma_{sca}/\Gamma_{rad}$. Because the contribution of phonon scattering to $\Gamma_{sca}$ should decrease with decreasing temperature while $\Gamma_{rad}$ should be largely independent of temperature. Figure 3a shows the temperature dependence of the reflection spectra near the A-exciton resonance. At 270 K, the A-exciton absorption at 1.59 eV has 32% contrast and a linewidth of approximately 40 meV. With decreasing temperature, the linewidth of the exciton absorption narrows nearly 7-fold from 40 meV to 6.1 meV (figure 3c).

The exciton resonance also shifts with temperature, which could affect the enhancement of the radiative decay rate due to the mirror. As shown in figure 3b, the temperature evolution of exciton energies can be fit well by the Varshni formula $E_g = E_0 - \alpha T^2/(T + \beta)$ that describes the bandgap increase with decreasing temperature[25]. Fitting yields $E_0 = 1.653$ eV, $\alpha = 0.48$ meV/K and $\beta = 211$ K, in good agreement with previously reported results[11]. Corresponding to the energy shift from 1.593 to 1.653 eV, we estimate by the transfer-matrix method that the local field factor changes by 12% (orange in figure 3b). It is a small change compared to the change of the total linewidth, and therefore the linewidth narrowing with temperature primarily results from reduced phonon scattering.

The temperature dependence of the normalized integrated absorption $A/A_0$ is shown in figure 3d. The gray curve shows the prediction from the perturbative theory where a linear relationship between the absorption and susceptibility is assumed, after accounting for the 12% change due to the shift of the exciton resonance. Clearly, the measured integrated absorption does not agree with the perturbative theory, showing instead the transition from the incoherent to coherent limit as $A/A_0$ decreases from near unity to a mere 33%. The large reduction of the integrated absorption upon cooling shows that the exciton-phonon scattering dominates at high temperatures, but it is greatly suppressed at low temperatures. As $\Gamma_{sca}$ is reduced, the system approaches the coherent limit.

The measured temperature dependence of $\alpha(T)$ can be compared with equation (1), which would also allow us to obtain $\Gamma_{rad}$ and $\Gamma_{sca}$. As shown in figure 3c, $\alpha(T)$ fits equation (1) very well. Here we use the measured linewidth $\hbar\Gamma_{tot}$, the temperature-independent $\hbar\Gamma_{rad}$ and $A_0$ are free parameters. The fit yields $\hbar\Gamma_{rad} = 4.0 \pm 0.2$ meV and $A_0 = 19.3 \pm 0.5$ meV. The radiative linewidth can also be checked independently as it is directly proportional to the measured integrated absorption at high temperatures, where $\Gamma_{sca} \gg \Gamma_{rad}$ and $A \approx A_0$. From the transfer matrix simulation, the $A_0$ value corresponds to $\hbar\Gamma_{rad} = 3.9 \pm 0.2$ meV, in excellent agreement with the fitted $\hbar\Gamma_{rad}$. The minimal normalized integrated absorption $A/A_0$ is measured to be 33% at 10K, which corresponds to a finite $\hbar\Gamma_{sca} = 2.1$ meV resulting from inhomogeneity and exciton-impurity scattering. It can be further reduced by improving the sample quality[26,27]. Nonetheless, the absorption decreases to below $A_0/2$ and should reach the critical coupling condition at $\Gamma_{sca} = \Gamma_{rad}$.

Figure 4a shows absorption spectra versus temperature from 110 K to 10 K. With decreasing temperature, the maximum absorption at the exciton resonance first increases, as expected from the reduction in linewidth at low temperature, but then decreases from 70 K to 10 K. A maximal absorption of 94.6% is achieved at $T_c = 70$ K, showing critical coupling between the photons and excitons. The critical temperature of T=70 K corresponds to the point at which $\Gamma_{rad} = \Gamma_{sca}$ and the integrated absorption $A/A_0 = 50\%$, as shown in figure 3c and d, respectively. Thus we have demonstrated that critical coupling between excitons and free-space photons is possible in a monolayer TMD.

Critical coupling phenomena have broad applications to light control, energy harvesting and optical sensing[12,14,19]. Perfect absorption can also be used in photodetectors with high quantum efficiency for quantum optics applications[12,13,24,28]. With an exceptionally large radiative linewidth, the monolayer TMD offers a highly flexible system for controlling the critical coupling. Besides temperature variation, we show two other tuning schemes that are easily accessible experimentally: tuning the $\Gamma_{sca}$ by laser excitation and tuning $\Gamma_{rad}$ by the distance from the mirror.

The strong exciton nonlinearity in TMDs enables tuning $\Gamma_{sca}$ using a pulsed laser. The pulsed laser has enough spectral bandwidth to cover the exciton peak for reflection measurements. As shown in figure 4b, with increasing excitation power, the absorption peak first increases in height then decreases, with a nearly perfect absorption of 99.6% where the critical coupling condition is met. At the same time, the linewidth of the absorption peak increases monotonically with increasing excitation power, consistent with the change in absorption resulting from changing $\Gamma_{sca}$. Solving equation (1) and maintaining the condition that $\Gamma_{total} = \Gamma_{sca} + \Gamma_{rad}$, we obtain $\Gamma_{sca}$ and $\Gamma_{rad}$, as shown in the upper panel of figure 4(b). As the excitation power increases by a factor of ten, the radiative linewidth remains constant while the scattering rate $\Gamma_{sca}$ quickly increases from 2 meV to 9 meV quadratically with power, as expected for exciton-exciton scattering. Our results demonstrate tuning of the absorption up to nearly 100% with a pulsed laser. This effect can be utilized for ultrafast light modulators and can be extended to room temperature by increasing $\Gamma_{rad}$ via nano-photonic structures[12,29].

At last, we demonstrate the possibility of achieving critical coupling by tuning $\Gamma_{rad}$ with a piezo-controlled movable mirror, as shown in the inset of figure 4c[30]. Tuning the spacing between the sample and mirror modifies the local photon density of states and in turn the radiative decay rate $\Gamma_{rad}$, while the scattering linewidth is unaffected. The absorption spectra at different mirror distances are taken at 50 K, as shown in figure 4c. The spectra show a strong modulation of the absorption depth. The critical coupling condition is achieved when the monolayer is positioned at the anti-node of the light, with a maximal absorption of 98.9%. The $\Gamma_{sca}$ and $\Gamma_{rad}$ obtained are shown in the upper panel of figure 4c. Such strong modulation of the radiative linewidth simply using a mirror results from the large oscillator strength and two-dimensional nature of monolayer TMDs. Consequently, critical coupling can be reached even with a low radiative enhancement factor.

In conclusion, we show that strong and coherent exciton-photon interactions are established in 2D MoSe$_2$ monolayer crystals, leading to the reduction of integrated absorption and striking demonstrations of critical coupling and perfect absorption in a monolayer. The observed 67% reduction of the total exciton absorption, as exciton-phonon scattering is reduced by temperature tuning, indicates that the system transitions from the incoherent to coherent limit. At low temperature, the absorption process needs to be considered in a non-perturbative fashion due to the coherent reemission from the monolayer. Analogous to critical coupling of emitters in high-quality cavities, as the scattering and radiative rates of the exciton equal, we achieve perfect absorption up to 99.6% in a sub-nanometer thick monolayer in front of a mirror. The critical coupling condition is readily established by tuning temperature, a pulsed excitation laser, and a movable mirror, which tune the exciton-phonon, exciton-exciton and exciton-photon coupling, respectively. These results demonstrate 2D TMD monolayers as a highly experimentally accessible system for rich phenomena resulting from coherent exciton-photon interaction, with wide ranging applications from light modulators to photodetector and coherent optical computing devices.

## Methods

### Sample preparation

The sample was prepared by mechanical exfoliation of bulk $MoSe_2$ (from 2D Semiconductors, USA) and hBN crystals[31]. Using the all-dry viscoelastic stamping technique[32], layers were first identified by optical contrast on polydimethylsiloxane(PDMS) and then deterministically transferred on the DBR or sapphire substrate to obtain the hBN/$MoSe_2$ monolayer/hBN stacking. The homemade DBR mirror has a stop band ranging from 670 to 810nm. It is made of 12 pairs of e-beam evaporated $SiO_2$ and $TiO_2$ layers on top of a sapphire substrate.

### Reflectance measurement

The fabricated samples are mounted in a closed-loop cryostat(Montana Instruments)for temperature dependence measurements. A broadband tungsten lamp or a femtosecond Ti:sapphire laser (with 20nm bandwidth) is focused onto the sample via a 50× long working distance objective lens with 0.42 numerical aperture, and the reflected light is selected by an aperture equivalent to a spot size ~2 μm in diameter. The light is collected into a grating spectrometer(Princeton Instruments)with a spectral resolution of 0.2 nm together with a charge-coupled device for recording light intensity. The reflectance spectrum is determined by $R_s/R_m(\omega)$ , where $R_s(\omega)$ and $R_m(\omega)$ are the reflected intensity from the mirror with and without the encapsulated $MoSe_2$.

### Acknowledgments

We gratefully thank Professor Mack Kira and Professor Bernhard Urbaszek for fruitful discussions. J.H. and H.D. acknowledge the support by the Army Research Office(ARO) under Grant # W911NF-17-1-0312(MURI: Room Temperature 2D Polaritronics with van der Waals Heterostructures). Y.-H. C. acknowledges the support by the Ministry of Science and Technology in Taiwan for the Postdoctoral

Research Abroad Program and Grant #106-2917-I-564-021. E.C. acknowledges a Labex NEXT travel grant.

**Author contributions**

J.H., E.M., H.D. and F.W. conceived the research idea. J.H. and E.M. carried out optical measurements and analyzed the data. E. C., Y.-H. C., T.-C. C., C.-Y.H. and M.-H. W. fabricated the samples. Theoretical investigation is done by J.H., E.M., S.C., F.W. and H.D.. J.H. and H.D. wrote the manuscript with inputs from all authors.

**Competing interests**

The authors declare no competing interests.

**Data availability**

The data that support the plots within this paper and other findings of this study are available from the corresponding author upon reasonable request.

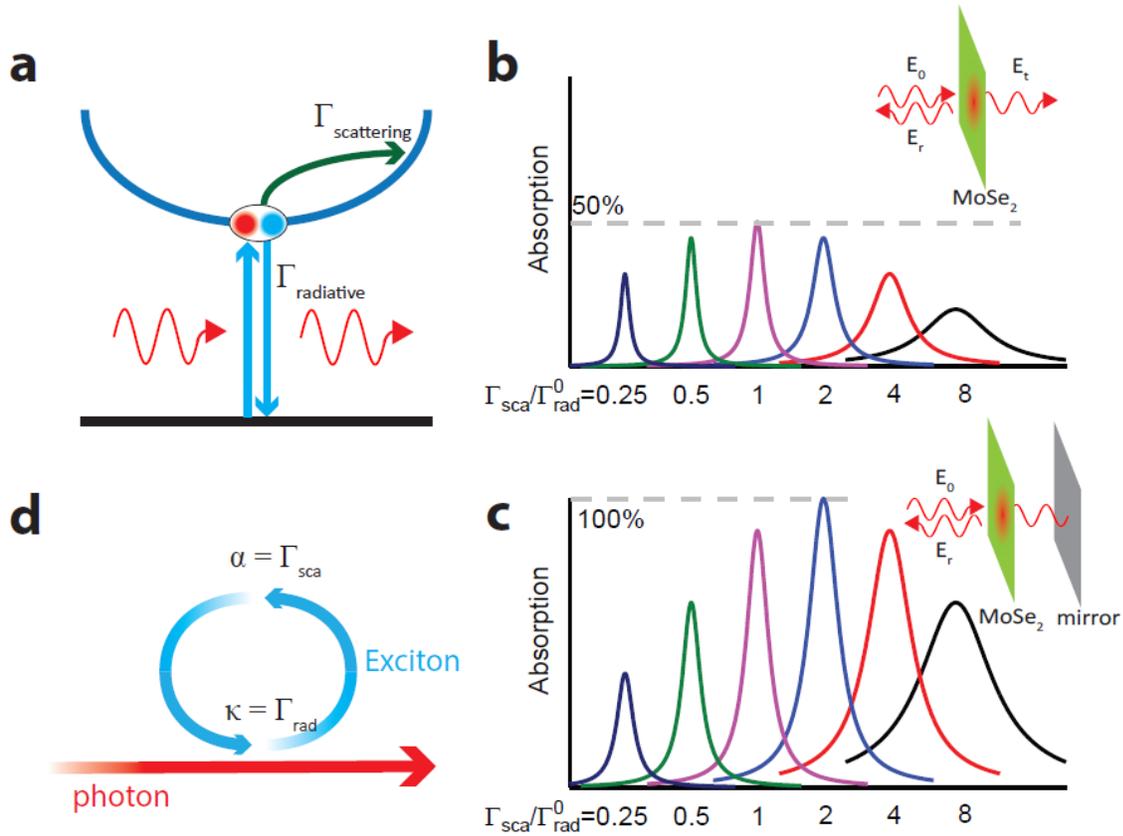

**Figure 1. Absorption mechanism in an excitonic system. a**, Microscopic mechanism of excitonic absorption. After excitation by absorbing a photon, the exciton polarization decays through either radiative recombination or scattering to other states, with rates $\Gamma_{rad}$ and $\Gamma_{sca}$, respectively. Only $\Gamma_{sca}$ causes energy loss and induces optical absorption. **b,c**, Illustration of absorption spectra of a free-standing 2D film(b) and a 2D film in front of a mirror(c) with varying scattering rates. In the free-standing 2D cases, the total absorption decreases as scattering processes are suppressed. The maximum absorption can reach 50% and 100% in the (b) and (c) case, respectively, when the integrated area is decreased to half. **d**, An analogy to a ring resonator for the case in (c). Critical coupling effect between free-space photons and excitons can occur in TMD monolayers. The photon channel is coupled with the exciton channel with an input coupling efficiency of $\Gamma_{rad}$. While the energy is transferred to the exciton channel, scattering processes can create loss with rate $\Gamma_{sca}$. The critical coupling condition is reached if $\Gamma_{sca} = \Gamma_{rad}$, and the absorption approaches 100%.

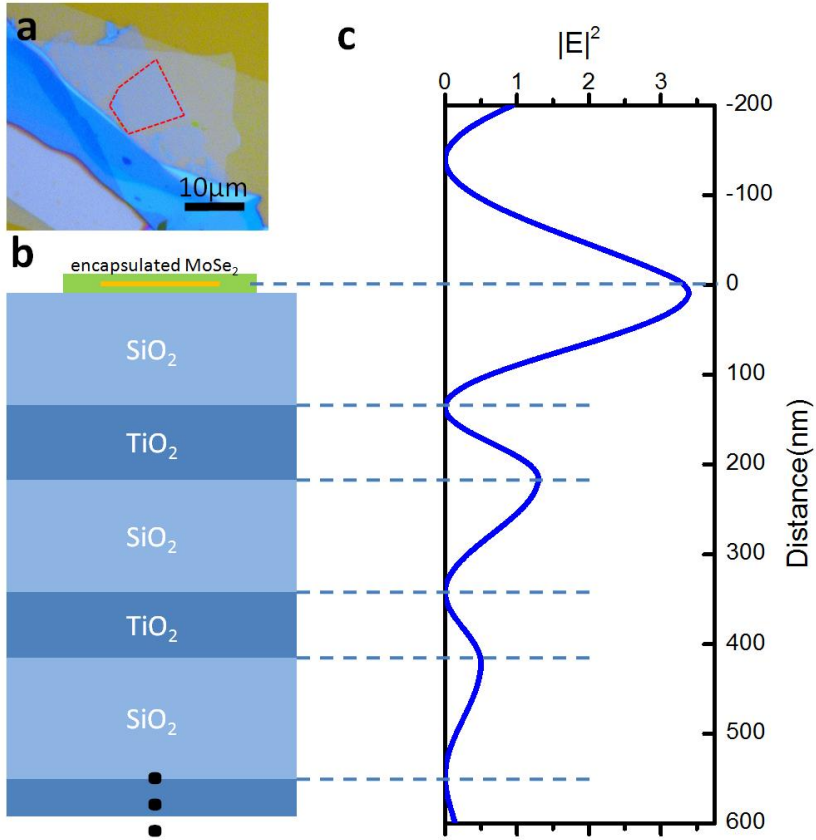

**Figure 2. Sample structure and local field enhancement from a distributed Bragg reflector(DBR).**
**a**, an optical microscope image of the sample. The red dotted lines encircle the MoSe$_2$ monolayer area. **b,** Schematic of the sample structure. The sample is designed such that MoSe$_2$ monolayer is λ/4 away from the DBR mirror to ensure the largest absorption and radiative enhancement. **c,** Simulation of the position-dependent squared electric field for photon energy 1.650eV, showing the MoSe$_2$ monolayer is close to the anti-node of the standing light wave.

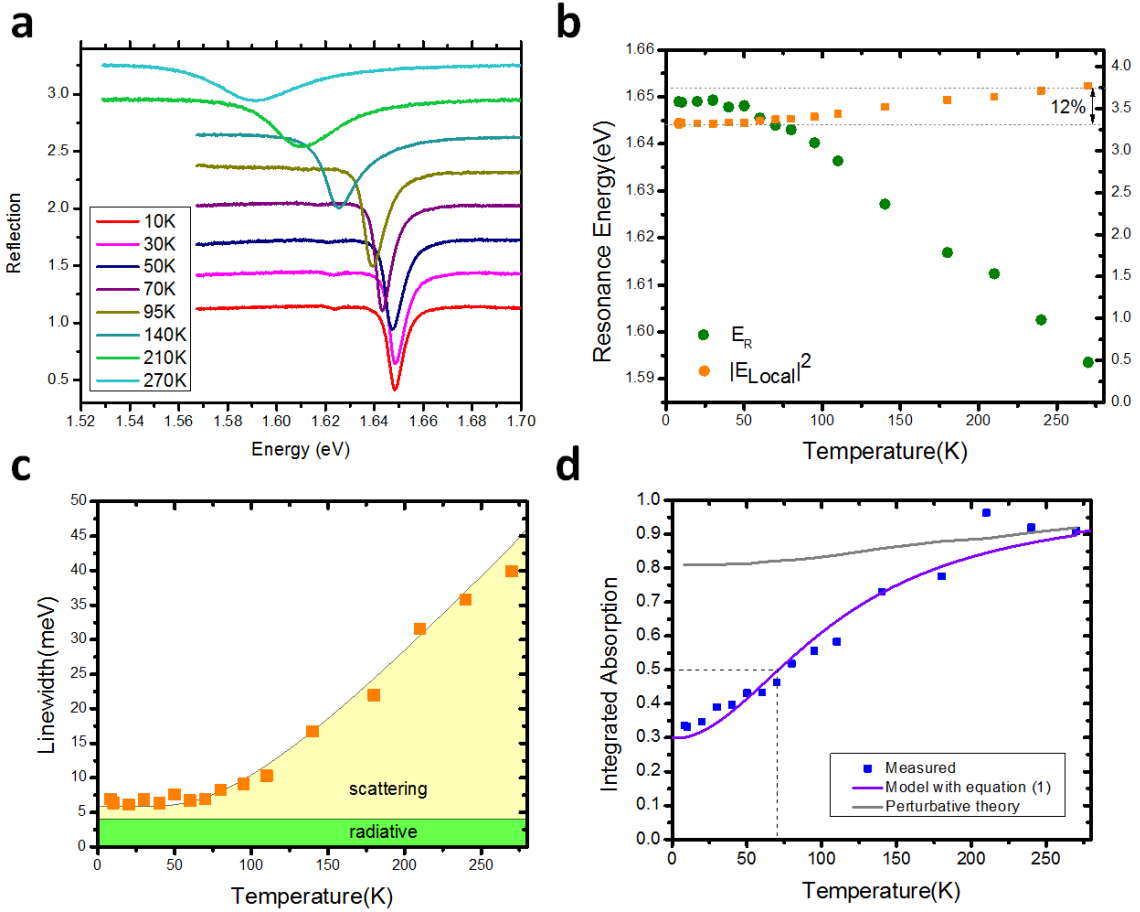

**Figure 3. Reduced optical absorption in the coherent limit. a**, Measured reflection spectra near the A-exciton resonance of the MoSe$_2$ monolayer at different temperatures. Spectra are offset for clarity. **b,** The temperature-dependent exciton resonance energy is shown in olive squares. The simulated $|E_{local}|^2$ at MoSe$_2$ position for the corresponding exciton wavelength at each temperature is plotted in orange squares. The absorption enhancement factor varies within 12% over the whole temperature range. **c,** Measured exciton linewidths versus temperature. The shaded areas show linewidth contributions from radiative(green) and scattering(yellow) broadenings. **d,** The measured(blue squares) and fitted(violet curve) integrated absorption as a function of temperature. The gray line shows the prediction from the perturbative theory taking into account the wavelength-dependent absorption enhancement. A remarkable 67% reduction in integrated absorption is observed experimentally, indicating coherent absorption dominates.

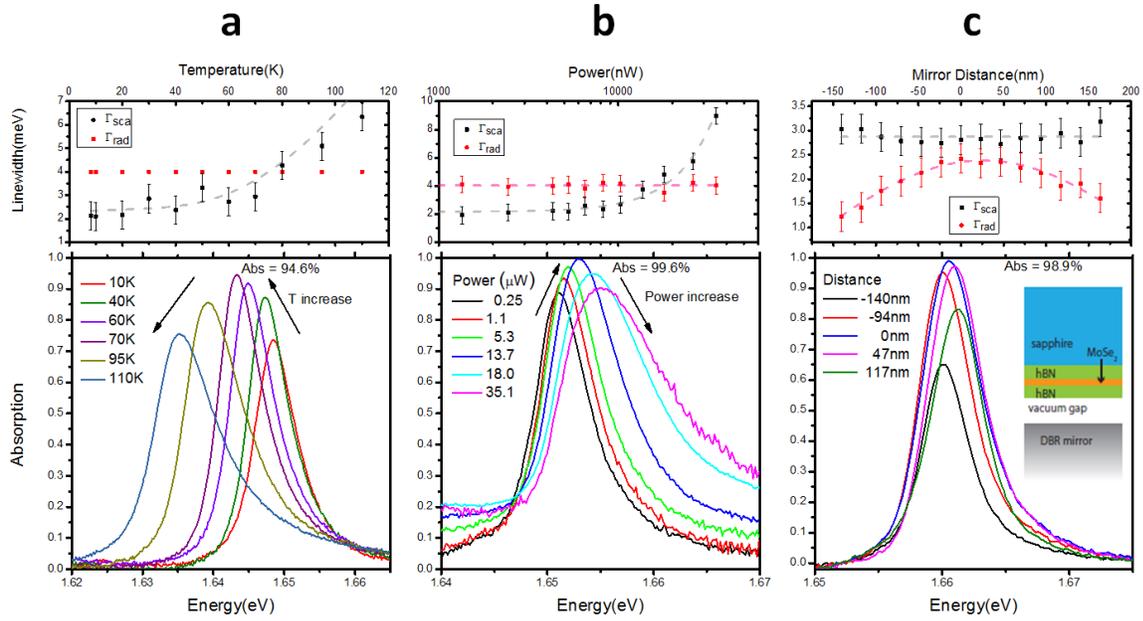

**Figure 4. Observation of critical coupling phenomenon between an exciton and a free-space photon. a-c**, Measured absorption spectra(lower panels) with temperature(a), laser intensity(b), and mirror distance(c) as tuning parameters. All three methods show clear critical coupling effect with nearly 100% maximum absorption of 94.6%, 99.6% and 98.9%, respectively. The extracted $\hbar\Gamma_{rad}$ and $\hbar\Gamma_{sca}$ are plotted in the corresponding upper panels. The critical coupling condition occurs when temperature, laser intensity and mirror distance are tuned such that $\Gamma_{rad} = \Gamma_{sca}$.